\documentclass[aps,prl,twocolumn,floatfix, showpacs,superscriptaddress]{revtex4-1}  % for review and submission

\usepackage[pdftex]{graphicx} %[dvips]
\usepackage{natbib}
\usepackage{dcolumn}   % needed for some tables
\usepackage{bm}        % for math
\usepackage{amssymb}
\usepackage{xcolor}    % for math
\usepackage{gensymb}
\usepackage[colorlinks=true,linkcolor=blue,citecolor=blue,urlcolor=blue]{hyperref}
\newcommand\phs{Pt$_2$HgSe$_3$}
\newcommand\ga{$\Gamma$}

\newcommand{\one}{Fig.~\ref{Fig1}}
\newcommand{\two}{Fig.~\ref{Fig2}}
\newcommand{\three}{Fig.~\ref{Fig3}}

\newif\ifshowcomments\showcommentstrue

\begin{document}

% The following information is for internal review, please remove them for submission
%\widetext
%\leftline{Version 01 as of \today}
%\leftline{Primary authors: Ir\`ene Cucchi}
%\leftline{To be submitted to PRL}
%\leftline{Comment to {\tt d0-run2eb-nnn@fnal.gov} by xxx, yyy}
%\centerline{\em D\O\ INTERNAL DOCUMENT -- NOT FOR PUBLIC DISTRIBUTION}

% the following line is for submission, including submission to the arXiv!!
%\hspace{5.2in} \mbox{Fermilab-Pub-04/xxx-E}

%\title{Observation of non-trivial surface states in the dual topological insulator \phs{}.}
\title{Bulk and surface electronic structure of the dual-topology semimetal Pt$_2$HgSe$_3$}

\author{I. Cucchi}
\affiliation{Department of Quantum Matter Physics, University of Geneva, 24 quai Ernest Ansermet, CH-1211 Geneva, Switzerland}
\author{A. Marrazzo}
\affiliation{Theory and Simulation of Materials (THEOS), and National Centre for Computational Design and Discovery of Novel Materials (MARVEL), \'Ecole Polytechnique F\'ed\'erale de Lausanne, CH-1015 Lausanne, Switzerland}
\author{E. Cappelli}
\affiliation{Department of Quantum Matter Physics, University of Geneva, 24 quai Ernest Ansermet, CH-1211 Geneva, Switzerland}
\author{S. Ricc\`o}
\affiliation{Department of Quantum Matter Physics, University of Geneva, 24 quai Ernest Ansermet, CH-1211 Geneva, Switzerland}
\author{F. Y. Bruno}
\affiliation{Department of Quantum Matter Physics, University of Geneva, 24 quai Ernest Ansermet, CH-1211 Geneva, Switzerland}
\affiliation{GFMC, Departamento de Física de Materiales, Universidad Complutense de Madrid, 28040 Madrid, Spain}
\author{S. Lisi}
\affiliation{Department of Quantum Matter Physics, University of Geneva, 24 quai Ernest Ansermet, CH-1211 Geneva, Switzerland}
\author{M. Hoesch}
\affiliation{Diamond Light Source, Harwell Campus, Didcot, United Kingdom}
\affiliation{Deutsches Elektronen-Synchrotron DESY, Photon Science, Hamburg 22607, Germany}
\author{T.K. Kim}
\affiliation{Diamond Light Source, Harwell Campus, Didcot, United Kingdom}
\author{C. Cacho}
\affiliation{Diamond Light Source, Harwell Campus, Didcot, United Kingdom}
\author{C. Besnard}
\affiliation{Department of Quantum Matter Physics, University of Geneva, 24 quai Ernest Ansermet, CH-1211 Geneva, Switzerland}
\author{E. Giannini}
\affiliation{Department of Quantum Matter Physics, University of Geneva, 24 quai Ernest Ansermet, CH-1211 Geneva, Switzerland}
\author{N. Marzari}
\affiliation{Theory and Simulation of Materials (THEOS), and National Centre for Computational Design and Discovery of Novel Materials (MARVEL), \'Ecole Polytechnique F\'ed\'erale de Lausanne, CH-1015 Lausanne, Switzerland}
\author{M. Gibertini}
\affiliation{Department of Quantum Matter Physics, University of Geneva, 24 quai Ernest Ansermet, CH-1211 Geneva, Switzerland}
\affiliation{Theory and Simulation of Materials (THEOS), and National Centre for Computational Design and Discovery of Novel Materials (MARVEL), \'Ecole Polytechnique F\'ed\'erale de Lausanne, CH-1015 Lausanne, Switzerland}
\author{F. Baumberger}
\affiliation{Department of Quantum Matter Physics, University of Geneva, 24 quai Ernest Ansermet, CH-1211 Geneva, Switzerland}
\affiliation{Swiss Light Source, Paul Scherrer Institute, CH-5232 Villigen, Switzerland}
\author{A. Tamai}
\affiliation{Department of Quantum Matter Physics, University of Geneva, 24 quai Ernest Ansermet, CH-1211 Geneva, Switzerland}
%\phone{+41 22 379 62 15}
\email{Anna.Tamai@unige.ch}
%\input author_list.tex       % D0 authors (remove the first 3 lines
                             % of this file prior to submission, they
                             % contain a time stamp for the authorlist)
                             % (includes institutions and visitors)
\date{\today}

% A PRL abstract should not exceed 600 characters
\begin{abstract}
We report high-resolution angle resolved photoemission measurements on single crystals of Pt$_2$HgSe$_3$ grown by high-pressure synthesis. Our data reveal a gapped Dirac nodal line whose (001)-projection separates the surface Brillouin zone in topological and trivial areas. In the non-trivial $k$-space range we find surface states with multiple saddle-points in the dispersion resulting in two van Hove singularities in the surface density of states. Based on density functional theory calculations, we identify these surface states as signatures of a topological crystalline state which coexists with a weak topological phase.
\end{abstract}

\maketitle

%A PRL article should not exceed 3750 words

%\section{\label{sec:level1}First-level heading}
% sections are not used for PRL papers

The prediction of the quantum spin Hall effect in graphene by Kane and Mele triggered a reformulation of band theory incorporating the concept of topology~\cite{kane2005, kane2005_z2}. Demonstrated first in HgTe/CdTe quantum well structures~\cite{bernevig_quantum_2006,Konig2007,Konig2010}, a quantum spin Hall state has subsequently been identified in exfoliated 1T' WTe$_2$~\cite{Tang2017a, Fei2017, Taniguchi2018, Cucchi2019} and was reported in bismuthene grown on SiC~\cite{Reis2017}. 
Recently, a robust quantum spin Hall insulator (QSHI) with a gap of up to 0.5~eV has also been predicted in monolayer (ML)  \phs ~\cite{Marrazzo2018}, raising the possibility of a highly insulating state up to room temperature in a van der Waals material.

When stacked to form a three-dimensional (3D) crystal, a 2D QSHI generically turns into a weak topological insulator (WTI) with no protected states on the top and bottom surface of the crystal \cite{Hasan2010}. Intriguingly though, recent theoretical work found a far richer scenario for bulk \phs. Refs.~\cite{Facio2019, Ghosh2019} predicted that \phs{} is one of only a few known dual topological insulators, and may host different surface states protected by symmetries that are unrelated to the QSHI state~\cite{Eschbach2017, Rusinov2016, Avraham2017, Zeugner2018}. 
Specifically, \phs{} was found to be a topological crystalline insulator (TCI) arising from a three-fold mirror symmetry in addition to a weak topological insulator protected by the preservation of translational symmetry in the stacking of the layers. 
However, to date little is known from experiment about the bulk band structure supporting the different topological phases of \phs~\cite{Kandrai2019} and the key-signature of the TCI, a topological surface state on the (001) surface, has not been reported.

Here we present an angle-resolved photoemission (ARPES) study of cleaved bulk \phs{} single crystals. We experimentally observe a spin-split surface state with multiple saddle point singularities on the (001) surface and show how this state emerges from a nodal line gapped by spin-orbit interactions.
Based on first principles calculations, we identify this surfaces state as the signature of a TCI phase. Our work thus provides strong evidence for the dual topology of \phs.

\begin{figure*}[t]

  \includegraphics[width=1\textwidth]{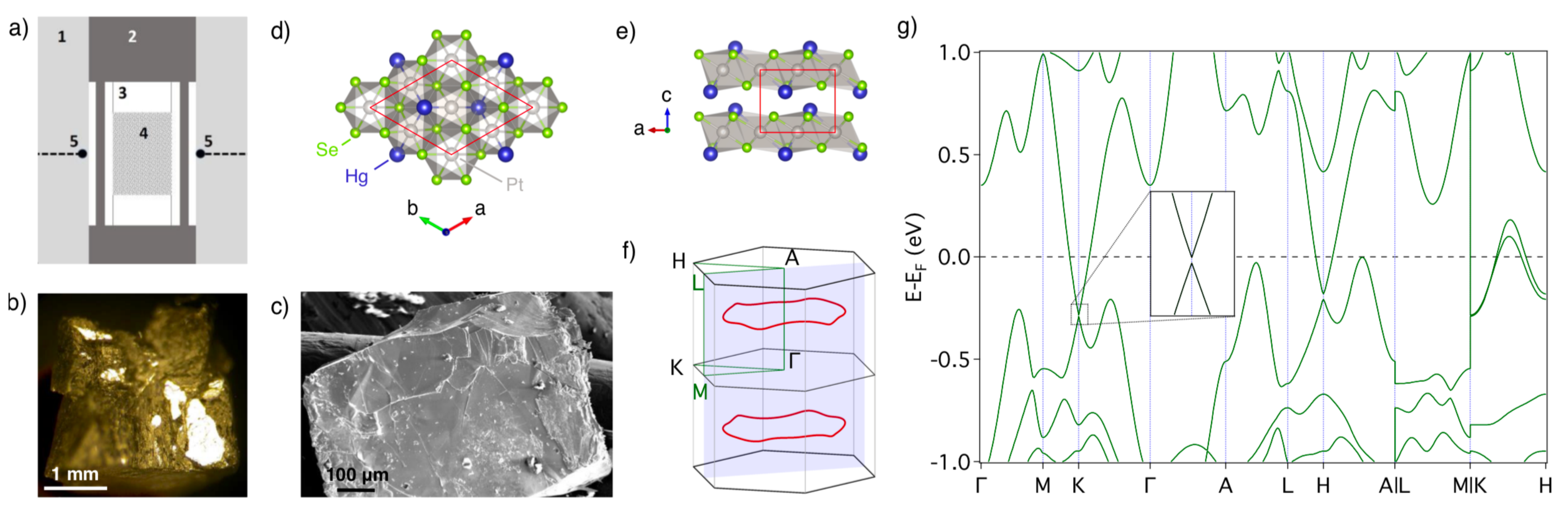}
    \caption{\label{Fig1}Bulk jacutingaite single crystals. a) Sketch of the high-pressure cell used to grow \phs {} single crystals: 1=pyrophillite cell; 2=graphite heater; 3=BN crucible; 4=sample; 5=thermocouples. (b) Broken as-grown boule: shiny crystal surfaces are visible inside the polycrystalline pellet. c) Scanning electron micrograph of a  \phs{} crystal extracted from the pellet. (d,e) Top and side views of the crystal structure. (f) Sketch of the 3D Brillouin zone. The blue shaded plane comprising the \ga, A, L and M points is one the three mirror planes relevant for the topology of the band structure. The nodal lines inside the BZ are marked in red.
    %The other two are obtained by a rotation of $\pm$ 60 degrees relative to the \ga A axis. 
    (g) DFT band structure calculation including SOC along the high symmetry directions. The inset shows an enlarged view of the Dirac-like dispersion at K, gapped by SOC interactions.}
\end{figure*}

First we describe the growth of jacutingaite \phs{} single crystals suitable for high-resolution ARPES experiments. Since Hg is highly volatile, we employed a high-pressure synthesis route using a high-temperature cubic-anvil press. Applying pressure in the GPa range prevents Hg from boiling off and allowed to stabilize a stoichiometric melt at temperatures as high as $900-1000^\circ$C. Elemental Pt (purity 99.98\%) and Se (purity 99.999\%), and the compound HgSe (purity 99.9\%) were thoroughly mixed with an Hg-exceeding nominal composition Pt$_2$Hg$_{1.1}$Se$_3$ and pelletized under Ar-atmosphere inside a glove box. The pellet (either $\sim0.14$~cm$^3$ or $\sim0.48$~cm$^3$, depending on the cell size) was inserted in a boron nitride crucible surrounded by a graphite heater that tightly fills the pyrophillite pressure cell  (\one(a)). The cell was first pressurized to $1.5-2$~GPa, then heated at a fast rate to the target temperature $850-1000^\circ$C, maintained at high temperature for $1-2$ hours then slowly cooled down ($50-75^\circ$C/h) to $650^\circ$C followed by a quench to room temperature. Crystalline platelets forming inside the solidified pellet could then be extracted mechanically (\one(b,c)). The crystal sizes obtained by this procedure were in the range of $0.5 - 1$~mm in lateral dimension and $50-100$~$\mu$m in thickness. 
Crystals were found to have a trigonal unit cell (space group \textit{P$\bar{3}$m1}) 
with lattice parameters $a = b = 7.3575(1)$~\AA{} and $c = 5.2902 (1)$~\AA, in good agreement with previously reported structural studies~\cite{Vym2012}. The chemical composition as measured by Energy Dispersive X-ray spectroscopy in a scanning electron microscope was \phs, within the uncertainty of the EDS probe. 

Synchrotron based ARPES measurements were carried out at beamline I05 of Diamond Light Source using photon energies of 19~eV - 80~eV~\cite{Hoesch2017}. Samples were cleaved \textit{in situ} along the $ab$-plane at temperatures $<20$~K. ARPES data were taken at $\sim6$~K with energy and momentum resolutions set to  $\sim 15$~meV and 0.02~\AA$^{-1}$, respectively.

Density-functional theory (DFT) calculations were performed with the Quantum ESPRESSO distribution \cite{giannozzi_quantum_2009,giannozzi_qe_2017}. We used a structure with the experimental lattice parameters reported above and relaxed the atomic positions using the non-local van der Waals functional vdW-DF2~\cite{lee_df2_09} with C09 exchange~\cite{cooper_c09_10}, with pseudopotentials from the SSSP precision library v1.0~\cite{prandini_precision_2018} (100~Ry of wavefunction cutoff and a dual of 8). The Brillouin zone was sampled with a $\mathbf{k}$-point density of 0.09~\AA$^{-1}$, corresponding to a $12\times12\times14$ Mokhorst-Pack grid, combined with a cold smearing~\cite{mv_smearing_99} of 0.015~Ry to smoothen the Fermi surface. Band structures and surface spectral densities were computed~\footnote{For these calculations we adopt Optimized Norm-Conserving Vanderbilt (ONCV~\cite{hamann_oncv_13}) pseudopotentials (either scalar or fully relativistic) from the PseudoDojo library \cite{dojo_paper_18} with 80~Ry of wavefunction cutoff and a dual of 4. Wannier functions are then obtained with the WANNIER90 \cite{mostofi_updated_2014} code using as initial projections $p$-orbitals on Se atoms, as well as $s$- and $d$-orbitals on both Pt and Hg atoms, and a $6\times6\times6$ $\mathbf{k}-$point grid. Spectral densities are then computed using Wannier Tools~\cite{wannier_tools_18}. Part of the calculations are powered by the AiiDA \cite{pizzi_aiida_16} materials' informatics infrastructure.} by mapping the electronic properties of \phs{} obtained with the PBE functional~\cite{perdew_pbe_96} onto maximally-localized Wannier functions~\cite{wannier_review_12}.

Bulk jacutingaite consists of layers of \phs{} stacked along the \textit{c} direction and held together mainly by van der Waals forces (\one(d,e)). Each layer comprises Pt atoms sandwiched between layers of Se and Hg atoms forming a buckled honeycomb structure. The natural cleavage plane of \phs{} is (001) with a Hg-terminated surface. The structure has an inversion symmetry around the center of the unit cell and three mirror planes perpendicular to the $ab$ plane, related to one another by a C$_3$ rotation. One of these mirror planes is indicated in light blue in the 3D Brillouin Zone (BZ) of \one(f).

The DFT bulk band structure computed including spin-orbit coupling (SOC) is displayed in \one(g). In agreement with previous calculations \cite{Facio2019, Ghosh2019} we find that \phs{} is a compensated semimetal with a band overlap of $\sim400$~meV.
At the K and H points one can readily identify Dirac-like cones which are gapped by SOC. In bulk \phs{} this gap is only a few meV, in sharp contrast with the large gap found in ML \phs{} at the equivalent point. This suppression is due to a strong interlayer coupling~\cite{Marrazzo_prep}, which is also responsible for the large band dispersion found along $k_z$.
The latter is particularly evident along the KH direction, where the bands forming the gapped Dirac cone cross the Fermi level twice, resulting in Fermi surface pockets of different carrier types along $k_z$. 
The nearly parallel dispersion of these two bands along KH further implies that the Dirac cones at K and H are connected by a  nodal line, mildly gapped by SOC.
Importantly, as emphasized in Ref.~\cite{Ghosh2019}, DFT calculations show 
a second nodal line forming a closed loop around the $\Gamma$A axis inside the BZ as indicated in \one(f) by red lines. In the following we show experimental evidence for the latter and characterize the topological surface state that develops from it.
%["Dirac cone" is line node too]

\begin{figure*}[t]

  \includegraphics[width=0.95\textwidth]{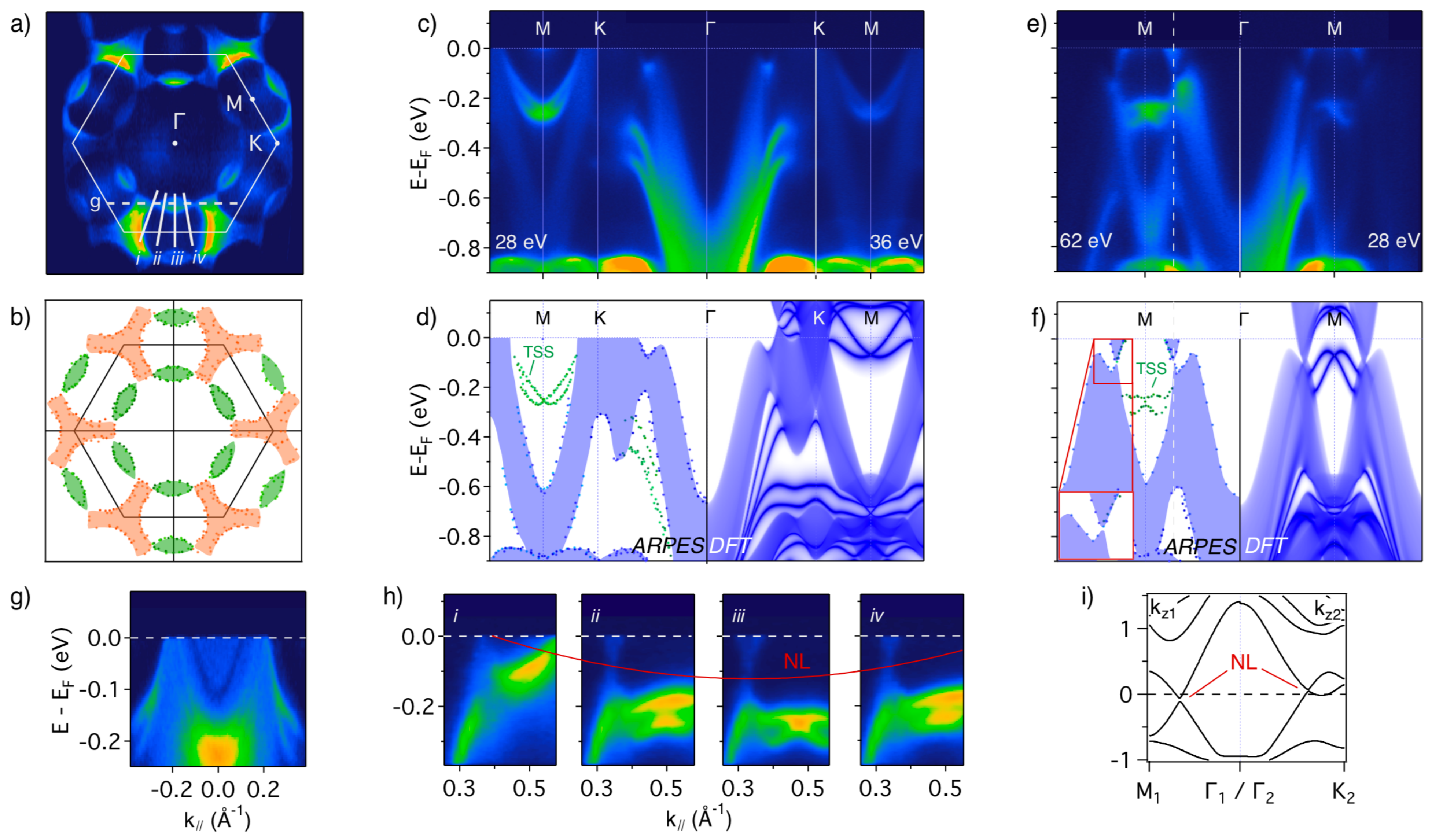}
    \caption{\label{Fig2} Bulk and surface electronic structure. a) ARPES Fermi surface measured with $h\nu=19$~eV excitation energy and $p$-polarization. b) Projected Fermi surface contours of hole (orange) and electron (green) pockets extracted from measurements at 19~eV and 62~eV photon energies. 
    c,e) ARPES dispersion plots measured along \ga K and \ga M with photon energies indicated in the figure.  d, f) Projected bulk band structure along \ga K and the \ga M obtained by combining data at 19, 28, 36 and 62~eV photon energies (left half). The dispersion of the surface state is indicated with green dots. DFT calculations of the surface spectral density (right half).  g) Cut along the white dashed line marked on the Fermi surface in panel (a). h) Radial cuts across the electron-like lens pocket measured at the positions marked by solid white lines in panel (a). 
    The dispersion of the nodal line is traced by a red line.
    i) DFT bands calculated at $k_{z1} =  -0.2962$~\AA$^{-1}$ (left) and $k_{z2} = -0.2735$~\AA$^{-1}$ (right) showing the nodal line crossings along \ga$_1$M$_1$ and \ga$_2$K$_2$.}
\end{figure*}

We start by presenting the Fermi surface of bulk \phs{} as measured with $h\nu=19$~eV on the (001) surface (\two(a)). 
These data show well defined lens-shaped contours on either side of the M points and three-fold symmetric features resembling fidget spinners around the K-points. Such a coexistence of one-dimensional (1D) contours and extended areas of high intensity with sharp boundaries is typical for UV photoemission from highly 3D systems with 2D surface states~\cite{Lindroos1996}.
It generally arises from surface photoemission interfering with direct transitions that are themselves broadened by the finite $k_z$ integration resulting from the strong damping of the final state~\cite{Lindroos1996,Miller1996}.
This can result in an almost complete $k_z$ integration of the spectral weight and thus in ARPES data closely resembling a surface projection of the bulk band structure rather than a specific cut through the 3D BZ.

In \two(b) we summarize the extension of the $k_z$ projected Fermi surface pockets determined from measurements at different photon energies.
Consistent with the semi-metallic nature predicted by DFT, we find small pockets of opposite carrier type. The 'fidget spinners' at the K-points spanning each of them $\sim 6.7\%$ BZ (orange shaded areas) are hole-like while the lens-shaped pockets with a volume of $\sim 1.5\%$ BZ (green shaded areas) that appear to connect the hole pockets are electron like. We stress that these areas represent the surface projected Fermi surface of a highly 3D system.
They are thus reminiscent of extremal orbits as they are measured in quantum oscillation experiments but cannot be used to deduce the degree of charge compensation. The latter is evident from the DFT dispersion along KH, predicting that the 'fidget spinner' hole pocket changes its polarity and becomes electron-like for $k_z$ values around K and H.

\begin{figure*}[t]
  \includegraphics[width=0.8\textwidth]{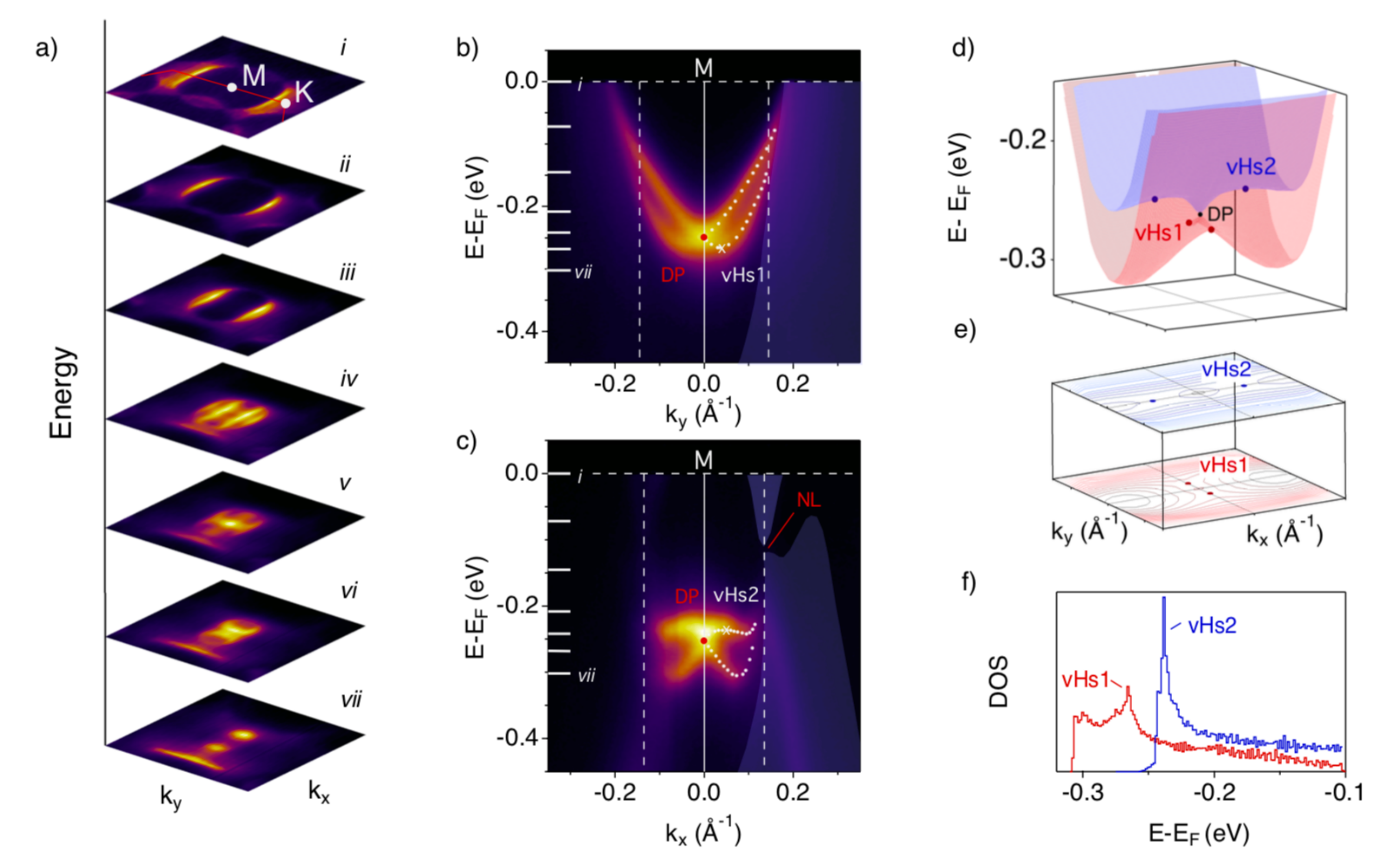}
    \caption{\label{Fig3} Dispersion of the topological surface state (TSS). a) Stack of constant energy maps around the M point. b,c) Dispersion along KMK $(k_y)$ and \ga M\ga $(k_x)$, respectively. The in-plane momentum on the horizontal axis is relative to the M point. On the right hand side the extracted surface and bulk band dispersions are superimposed on the ARPES data. d) 3D plot of the energy and momentum dispersion of TSS extracted over the momentum range centered at M indicated by vertical dashed lines in (b,c). e) Associated iso-energy contours of the lower (blue) and upper (red) surface state branches. f) Histograms of the corresponding $\epsilon(k)$ values showing sharp peaks at the energy positions of the van Hove singularities.}
\end{figure*}

Representative dispersion plots measured along the high symmetry directions \ga K and \ga M are shown in \two(c,e). 
As on the Fermi surface, the data show a coexistence of sharp 2D bands and broad features, which arise from a surface projection of bands with $k_z$ dispersion.
Evaluating the edges of the latter in dispersion plots acquired at different photon energies, we obtain the projected bulk band structure displayed as shaded blue areas on the left of \two(d,f). These areas are in good agreement with a DFT calculation of the momentum resolved surface density of states along the same path (\two(d,f) right).
Both experiment and theory show pronounced minima of the valence band at \ga{} and M and find that the valence band reaches a local maximum just below $E_F$  before crossing the Fermi level in the vicinity of K. An equally good agreement is observed for several other features that can be identified in experiments.
This confirms the interpretation of the ARPES intensity distribution as a bulk projection of the electronic structure and implies a good overall agreement of the full 3D electronic structure of \phs{} with DFT predictions.

%%%%%%%%%%%%%%%%%%%%%%%%%%%%%%%%%%%%%%%%%%%%%%%%%%%%%
%[revise this paragraph]
We now focus on the nodal line in the bulk electronic structure which is particularly important for the topological character of \phs. Moving from \ga{} towards M, our data show that conduction and valence bands nearly touch forming Dirac-like cones at the positions marked by dashed vertical lines in \two(e,f) and highlighted in the inset of \two(f). 
The radial cuts of this area displayed in \two(h) 
show that the Dirac cone moves in energy but does not noticeably change its small gap.
It thus forms a narrowly gapped nodal line which disperses across the Fermi level and forms a loop around \ga A. 
In \two(g) we directly visualize the dispersion of the gapped nodal line by 
extracting a cut  along the long axis of the lens-like electron pocket (dashed line in \two(a)).
This shows a sharp suppression of intensity in the surface projected spectral weight along a nearly parabolic contour (dark blue in the false color plot) that can be identified with the gap along the nodal line. The width of this suppression suggest a gap of $\sim 25$~meV at the bottom of the electron pocket, consistent with the SOC gaps found by DFT. 

To illustrate the relevance of this nodal line we show in \two(i) band structure calculations along \ga M and \ga K at the $k_z$ values with minimal gap. From these calculations, it is evident that, the (001)-projection of the nodal line separates the surface Brillouin zone in an inner region with trivial band ordering and an outer region comprising the M and K points which displays a band inversion. Direct calculations show~\cite{Marrazzo_prep} that this band inversion leads to a non-trivial Zak phase~\cite{zak_1989}, thus implying the existence of surface states connecting valence and conduction band in the outer region of the Brillouin zone.
%%%%%%%%%%%%%%%%%%%%%%%%%%%%%%%%%%%%%%%%%%%%%%%%%%%%
%
 
Inside the projected bulk band gap centered at M we find prominent states with 2D character that are degenerate at the time reversal invariant M points but split in pairs away from M before merging into the bulk continuum in the vicinity of the gapped nodal line. These characteristics are hallmarks of surface states.
Our calculations shown on the right hand side of \two(d,f) reproduce the surface states qualitatively and confirm their topological character identified in earlier theoretical work~\cite{Facio2019, Ghosh2019, Marrazzo_prep}. We note, that such calculations do not include surface effects such as structural relaxations and a possible charge accumulation or depression,  and are thus not expected to be in quantitative agreement with experiment. We also find that the exact dispersion of surface states is sensitive to details of the simulations, including the number of bands mapped into Wannier functions. 

%We now discuss the origin and topological character of this surface state.
The observation of surface states on the (001) surface of bulk \phs{} is \textit{a priori} unexpected given that \phs{} can be described as a stack of 2D QSHIs that preserves translational symmetry and  thus supports a weak topological phase. Calculations of the 3D $\mathbb{Z}_{2}$ topological invariants indeed find $(0; 001)$ \cite{Vergniory2019, Facio2019, Ghosh2019}, confirming the weak topological insulator phase generally characterized by gapless modes on the lateral surfaces
but fully gapped states on the top and bottom surfaces.
Clearly the surface states found on the (001) surface do not fit this description and thus must be the manifestation of a different topological phase.
This phase has been identified in Refs.~\cite{Facio2019, Ghosh2019} as a topological crystalline insulator (TCI), with a non-zero mirror Chern number ($|C_m|=2$) associated with the three-fold mirror symmetry of the crystal. In such a case, topologically protected surface states are expected on crystal surfaces that preserve the mirror symmetry \cite{Hsieh2012}, which is the case for the cleaved (001) surface. 
%{\color{red}The next sentence should state more clearly why this SS is a consequence of the TCI phase.} 
The surface state observed by ARPES at the M point can be identified as the signature of such a TCI phase, providing strong experimental evidence for the predicted dual topology of \phs.
We note that \phs{} has also been classified among the higher-order topological insulator with the single $\mathbb{Z}_{4}$ indicator equal to 2 \cite{Schindler2018, Vergniory2019}. Hence, topological surface states are expected not only on lateral, top and bottom surface but might also exist along hinges. A direct experimental confirmation of the hinge states on \phs{} will be difficult though. 
%owing to the metallic character of both the surfaces and the bulk.

We finally discuss the peculiar dispersion of the TSS.
\three(a) shows a stack of constant energy maps of the TSS centered at the M points. Moving from the Fermi level towards higher binding energies we observe an abrupt change in the distribution of its spectral intensity as the Dirac point is crossed. At low binding energies the TSS forms pairs of croissant-shaped contours elongated along $k_x$ (panels i-iii), whereas at high binding energies the constant energy contours become nearly elliptical with the long axes parallel to $k_y$ (panel vi).
A non-trivial dispersion is also evident in the two perpendicular cuts shown in \three(b,c).  Along $k_y$ one recognizes the characteristic dispersion of a Rashba-like spin split surface state.
Along $k_x$, however, the upper branch of the TSS has a non-monotonic dispersion with shallow maxima away from the Dirac point.

To analyze the surface density of states resulting from this unusual dispersion, we first
%A 3D plot of the energy and momentum dispersion of TSS is displayed in \three(d). It was obtained by 
fit the energy position of the two surface state branches in a series of dispersion cuts measured along $k_y$  for 20 different values of $k_x$, spanning a region of $\pm$ 0.12 \AA$^{-1}$ around M.
From these fits, we then obtain the hyper-surfaces of the energy-momentum dispersion displayed in \three(d,e).
This analysis demonstrates that in contrast to the TSSs found in conventional topological insulators \cite{Xia2009, Chen2009}, the TSS on \phs{} has a complex shape with multiple saddle points resulting in two van Hove singularities in the density of states. 
The latter can clearly be identified  in the form of sharp peaks at -266 meV and -239 meV in the histogram of the $\epsilon(k)$ values for the lower and upper branch, respectively.
Such singularities were predicted as generic features of TSSs emerging from narrowly gapped line nodes~\cite{Singh2018, Ghosh2019} but have not been reported before.
Tuning these singularities to the Fermi level by bulk or surface doping should favor ferro- or metamagnetic instabilities~\cite{Liu2017,Binz2004} and was recently predicted to stabilize topological superconductivity~\cite{Yuan2019, Wu2018}. %\textcolor{red}{[refs.??]}.

In conclusion, we reported the synthesis of jacutingaite \phs{} single crystals and demonstrated their dual topological nature by comparing the bulk and surface electronic structure determined from ARPES experiments with DFT calculations.
The good overall agreement with these calculations further supports the recent prediction of a robust QSHI phase in monolayer \phs~\cite{Marrazzo2018}.

We gratefully acknowledge discussions with A. Kuzmenko, A. Morpurgo, C. Renner and F. Schindler.
The experimental work was supported by the Swiss National Science Foundation (SNSF).
A.M and N.M. acknowledge support by the NCCR MARVEL of the SNSF. M.G.\ was  supported by the SNSF through the Ambizione program. Simulation time was awarded by CSCS on Piz Daint (production projects s825 and s917) and by PRACE on Marconi at Cineca, Italy (project id.\ 2016163963). We acknowledge Diamond Light Source for time on beamline I05 under Proposal No. SI18952.

%\bibliographystyle{plain}
%\bibliography{PtHS.bib}

%merlin.mbs apsrev4-1.bst 2010-07-25 4.21a (PWD, AO, DPC) hacked
%Control: key (0)
%Control: author (8) initials jnrlst
%Control: editor formatted (1) identically to author
%Control: production of article title (-1) disabled
%Control: page (0) single
%Control: year (1) truncated
%Control: production of eprint (0) enabled
%

\end{document}